\documentclass[fleqn,usenatbib]{mnras}

\usepackage[T1]{fontenc}
\usepackage{ae,aecompl}

\usepackage{graphicx}    
\usepackage{amsmath}    
\usepackage{amssymb}    
\usepackage{stfloats}
\usepackage{bm}
\usepackage[dvipsnames]{xcolor}  
\usepackage{soul}
\usepackage{tabulary,booktabs}

\setulcolor{blue}   
\setstcolor{red}  
\sethlcolor{green}  

\title[$H_0$ with L.O.S. halos]{The impact of line-of-sight structures on measuring $H_0$ with strong lensing time-delays}

\author[Nan Li et al.]{
Nan Li$^{1,2}$\thanks{E-mail: nan.li@nottingham.ac.uk},
Christoph Becker$^{3}$,
Simon Dye$^{1}$
\\
$^{1}$School of Physics and Astronomy, University of Nottingham, University Park, Nottingham, NG7 2RD, UK.\\
$^{2}$National Astronomical Observatories, Chinese Academy of Sciences, A20 Datun Road, Beijing 100012, China.\\
$^{3}$Institute for Computational Cosmology, Department of Physics, Durham University, South Road, Durham DH1 3LE, UK.\\
}

\date{Accepted XXX. Received YYY; in original form ZZZ}

\pubyear{20**}

\begin{document}
\label{firstpage}
\pagerange{\pageref{firstpage}--\pageref{lastpage}}
\maketitle
\begin{abstract}
Measurements of The Hubble-Lemaitre constant from early- and local-universe observations show a significant discrepancy.  In an attempt to understand the origin of this mismatch, independent techniques to measure $H_0$ are required. One such technique, strong lensing time delays, is set to become a leading contender amongst the myriad methods due to forthcoming large strong lens samples. It is therefore critical to understand the systematic effects inherent in this method. In this paper, we quantify the influence of additional structures along the line-of-sight by adopting realistic light cones derived from the \textit{CosmoDC2}  semi-analytical extra-galactic catalogue. Using multiple lens plane ray-tracing to create a set of simulated strong lensing systems, we have investigated the impact of line-of-sight structures on time-delay measurements and in turn, on the inferred value of $H_0$. We have also tested the reliability of existing procedures for correcting for line-of-sight effects. We find that if the integrated contribution of the line-of-sight structures is close to a uniform mass sheet, the bias in $H_0$ can be adequately corrected by including a constant external convergence $\kappa_{\rm ext}$ in the lens model. However, for realistic line-of-sight structures comprising many galaxies at different redshifts, this simple correction over-estimates the bias by an amount that depends linearly on the median external convergence. We therefore conclude that lens modelling must incorporate multiple lens planes to account for line-of-sight structures for accurate and precise inference of $H_0$.
\end{abstract}


\section{Introduction}
\label{sec:intro}
The Hubble-Lemaitre constant, $H_0$, is a cornerstone of the standard cosmological model, setting the distance scale, age and critical density of the Universe. Accurate estimation of the value of $H_0$ is therefore critical for constraining cosmological models in the era of precision cosmology. However, presently, there is a significant mismatch between $H_0$ determined from early- and late-universe probes \citep{Riess2019b,Verde2019}, for instance, measurements of the Cosmic Microwave Background \citep[CMB; see][]{WMAP9,Planck2018} and Baryon Acoustic Oscillations \citep[BAO; see][]{Addison2018,DESY1H0-2020} and those made in the more local Universe using supernovae \citep[SNe;see][]{Dhawan2018,Macaulay2019}, the tip of the red giant branch \citep[TRGB; see][]{Freedman2019,Yuan2019} and Cepheid variables \citep{Riess2019,Pietrzynski2019}. Independent from any of the aforementioned methods, strong lensing time delays provide valuable measurements of $H_0$ \citep[e.g.,][]{Wong2019,Shajib2019} which may assist in the understanding of these discrepancies once systematic uncertainties in the technique are fully calibrated. With such systematics in mind, in this paper we focus on the effects of line-of-sight structure, one of the most dominant sources of error in the lens time delay method.

Strong lensing time delays are observed when a variation in flux of a strongly-lensed background source such as a quasar, supernova or a gravitational wave event is detected at different times between its multiple images. The deflection of the light path from the source due to the gravitational potential of a lens, as well as the structures along the line-of-sight, leads to both a geometrical and a gravitational delay of the arrival time of the light from the source. The geometrical delays are sensitive to $H_0$ \citep[see][]{Schneider1992}. Therefore, measuring the time delays and reconstructing the mass distribution of the lens accurately allows $H_0$ to be estimated. The existing relative paucity of strong-lens systems suitable for this method and the necessary long monitoring campaigns has somewhat limited the use of this technique but good progress has already been made with only a handful of systems \citep[e.g.,][]{Suyu2010,Suyu2013,Birrer2016,Wong2017,Bonvin2017,Wong2019,Chen2019,DAgostino2020}. However, this is set to dramatically change \citep{Oguri2010,Collett2015} with the advent of the
Rubin Observatory Legacy Survey of Space and Time\footnote{\url{https://www.lsst.org/}} (LSST), which will give rise to about 400 well-measured time delay systems to constrain $H_{0}$ to within only a few percent \citep{Liao:2014cka,Dobler2015}.

Even with precise time delay measurements, the reliability of estimates of $H_0$ depends on how faithfully the lens mass model follows the true lensing mass. Degeneracies and inadequacies in the parameterisation of the lens mass model can directly propagate into the inferred value of $H_0$ \citep[e.g., see][]{Scheinder2013,Sereno2014,XuD2016,Munoz2017,Tie2018,Tagore2018,Wertz2018, Wagner2018} as can selection effects within the lens sample \citep[see][]{Collett2016}. In addition, perturbative effects from sub-structure within the main lens and from structure along the line-of-sight can significantly modify time delays which can bias measurements of $H_0$ if not properly taken into account. One approach to account for these effects is to directly characterise perturbing structures identified in observations \citep[e.g.,][]{Wong2011,Momcheva2015,Rusu2017,Sluse2017,Wong2018}. Another common technique is to use external shear, $\gamma_{\rm ext}$, and external convergence, $\kappa_{\rm ext}$, in the lens model. By connecting cosmological simulations and real observations, an estimate of the distribution function of the amplitude of these external lensing effects can be obtained \citep[e.g.,][]{Suyu2010,Suyu2013,Greene2013,Collett2013,Rusu2017,Birrer2017,Tihhonova2018}. However, the corrections provided by $\gamma_{\rm ext}$ and $\kappa_{\rm ext}$ are isotropic and cannot properly capture the complexity of real perturbing structures. Motivated by this, more sophisticated approaches have been developed using multiple lens planes or approximations thereof \citep[e.g.,][]{McCully2014,Birrer2017,McCully2017}.

In this work, we investigate the influence of halos along the line-of-sight on measurements of $H_{0}$ by using multiple lens plane ray-tracing simulations. To obtain simulated time delays we construct the light cone of each lens from a state-of-the-art semi-analytic model \citep[CosmoDC2\footnote{\url{https://portal.nersc.gov/project/lsst/cosmoDC2}}; ][]{CosmoDC2-2019} based upon the large Outer Rim cosmological N-body simulation \citep{Heitmann2019}. By modelling these time delays with the same methods used for real data, we directly assess the biases introduced by line-of-sight effects and the efficacy with which these can be accounted for using external corrections such as $\gamma_{\rm ext}$ and $\kappa_{\rm ext}$.

The paper is structured as follows. We outline the methodology used for determining strong lensing time delays in the cases of the single-lens plane and multiple-lens planes in Section~\ref{sec:theory}. Details of the simulations and the process of estimating $H_0$ from the simulated data are given in Sections~\ref{sec:sims} and~\ref{sec:lens-modelling} respectively. We present our findings in Section~\ref{sec:results}, then conclude with a summary and discussion in Section~\ref{sec:conclusion}. The cosmological model adopted in this paper is that used by CosmoDC2: $\Lambda$CDM with $\Omega_{\Lambda} = 0.735$, $\Omega_{\rm M} = 0.265$, and $H_0 = 71$\,km\,s$^{-1}$\,Mpc$^{-1}$.


\section{Strong Lensing Time Delays}
\label{sec:theory}
In this section, we present a basic description of the theory of time-delays in strong lensing systems with multiply-lensed point sources we have used in this work, for the cases of single and multiple lens planes. Throughout the paper, we have applied the thin lens approximation. For more details, we refer the reader to \cite{Schneider1999} and \cite{Narayan1996}.

\subsection{Time Delays in Single Lens Planes}
\label{sec:theory-spl}
For the case of a lensing system with a single deflector, adhering to the thin lens approximation, one can project the three-dimensional mass distribution to a two-dimensional mass sheet normal to the line-of-sight from the observer to the source. The dimensionless surface mass density of a thin lens plane can be written as a function of the lens plane angular position vector, $\bm{\theta}$, as
\begin{equation}
\label{eq:kappa}
 \kappa(\bm{\theta}) = \Sigma(\bm{\theta} D_{\rm d})/\Sigma_{\rm crit}\,\,,
\end{equation}
with the critical surface mass density
\begin{equation}
\label{eq:sigma_crit}
    \Sigma_{\rm crit}=\frac{c^2}{4\pi G} \frac{D_{\rm s}}{D_{\rm d}D_{\rm ds}}\,\,,
\end{equation}
where $D_{\rm s}$ and $D_{\rm d}$ are the angular diameter distances from the source and lens to the observer respectively,
$D_{\rm ds}$ is the angular diameter distance from the lens to the source,
and $\Sigma(\bm{\theta} D_{\rm d})$ is the surface mass density of the lens. The lensing potential is given by
\begin{equation}
 \psi(\bm{\theta}) = \frac{1}{\pi}\int {\rm d}^2\bm{\theta^{'}} \kappa(\bm{\theta^{'}}) {\rm ln}|\bm{\theta}-\bm{\theta^{'}}| \,\,,
 \label{eq:psi_kappa}
\end{equation}
and the deflection angle vector is given by
\begin{equation}
 \bm{\alpha}(\bm{\theta})=\frac{1}{\pi} \int {\rm d}^2 \bm{\theta^{'}} \kappa(\bm{\theta^{'}}) \frac{\bm{\theta} -\bm{\theta^{'}}}{|\bm{\theta}-\bm{\theta^{'}}|^2}\,\,.
 \label{eq:alpha_kappa}
\end{equation}

Once the deflection field at the lens plane is known,
we can construct the lensing equation for a given set of source planes. For example, in the case of a single lens plane and a single source plane, the lensing equation is simply
\begin{equation}
 \bm{\beta}=\bm{\theta}-\bm{\alpha}(\bm{\theta})\,\,,
 \label{eq:lensing_eq_slp}
\end{equation}
where $\bm{\beta}$ is the angular source plane position vector that maps to $\bm{\theta}$ in the image plane (or, equivalently, ``lens plane'' for the case of single lens-plane). Based on Eq.~\ref{eq:lensing_eq_slp}, ray-tracing simulations can be performed from the observer, crossing the lens plane to the source plane to produce lensed images. For extended source-like galaxies, to create distorted lensed images, interpolation can be used in the source plane to map spatially varying surface brightness back to the image plane. However, for the point sources used in this work, one has to adopt triangle mapping and a barycentric coordinate system to solve the lensing equations numerically. Details of the approach are discussed in Sec.~\ref{sec:image-finding}.

In the case of a single lens plane, the delay of the arrival time of a light ray from the source to the observer is
\begin{equation}
    \label{eq:arrival_time}
    \tau(\bm{\theta}, \bm{\beta}) = \frac{(1+z_{\rm d})}{c} \frac{D_{\rm d} D_{\rm s}}{D_{\rm ds}} \left[ \frac{(\bm{\theta}-\bm{\beta})^2}{2} - \psi (\bm{\theta}) \right] \, ,
\end{equation}
where $z_{\rm d}$ is the redshift of the lens. The last term in Eq.~\ref{eq:arrival_time} is also known as the Fermat potential,
\begin{equation}
    \label{eq:phmt-pot}
    \Phi(\bm{\theta}, \bm{\beta}) \equiv \left[ \frac{(\bm{\theta}-\bm{\beta})^2}{2} - \psi (\bm{\theta}) \right]\,\,.
\end{equation}
This delay is undetectable, the true observable being the difference between the arrival time of two separate lensed images (say, image A and image B), $t_{\rm AB} \equiv \tau_{\rm A} - \tau_{\rm B}$. From Eq.~\ref{eq:arrival_time}, the time difference can be written
\begin{equation}
    \label{eq:time-daley-ab}
    t_{\rm AB} = \frac{D_{\Delta \tau}}{c} \Delta\Phi_{\rm AB}\,\,,
\end{equation}
where,
\begin{equation}
    \label{eq:d-delta-tau}
    D_{\Delta \tau} \equiv (1 + z_{\rm d}) \frac{D_{\rm d} D_{\rm s}}{D_{\rm ds}}
\end{equation}
and
\begin{equation}
    \label{eq:delta-phi}
    \Delta \Phi \equiv \Phi(\bm{\theta}_{\rm A}, \bm{\beta}) - \Phi(\bm{\theta}_{\rm B}, \bm{\beta})\,\,.
\end{equation}
Note that
\begin{equation}
    \label{eq:daz}
    D_a(z) = \frac{c}{H_0 (1+z)} \int^{z}_{0} \frac{d z^{'}}{E(z^{'})}
\end{equation}
where
\begin{equation}
    \label{eq:ez}
    E(z) = \sqrt{\Omega_r (1+z)^4 + \Omega_m (1+z)^3 + \Omega_k (1+z)^2 + \Omega_\Lambda}\,\, .
\end{equation}
These equations show that
\begin{equation}
    \label{eq:td-h0}
    t_{\rm AB} \propto D_{\Delta \tau} \propto \frac{1}{H_{0}}
\end{equation}
and thus $H_0$ can be measured from $t_{\rm AB}$ if the mass distribution of the lens is reconstructed accurately.

\subsection{Time Delays in Multiple Lens Planes}
\label{sec:theory-mpl}
\begin{figure*}
    \centering
    \includegraphics[width=0.9\textwidth, angle=0]{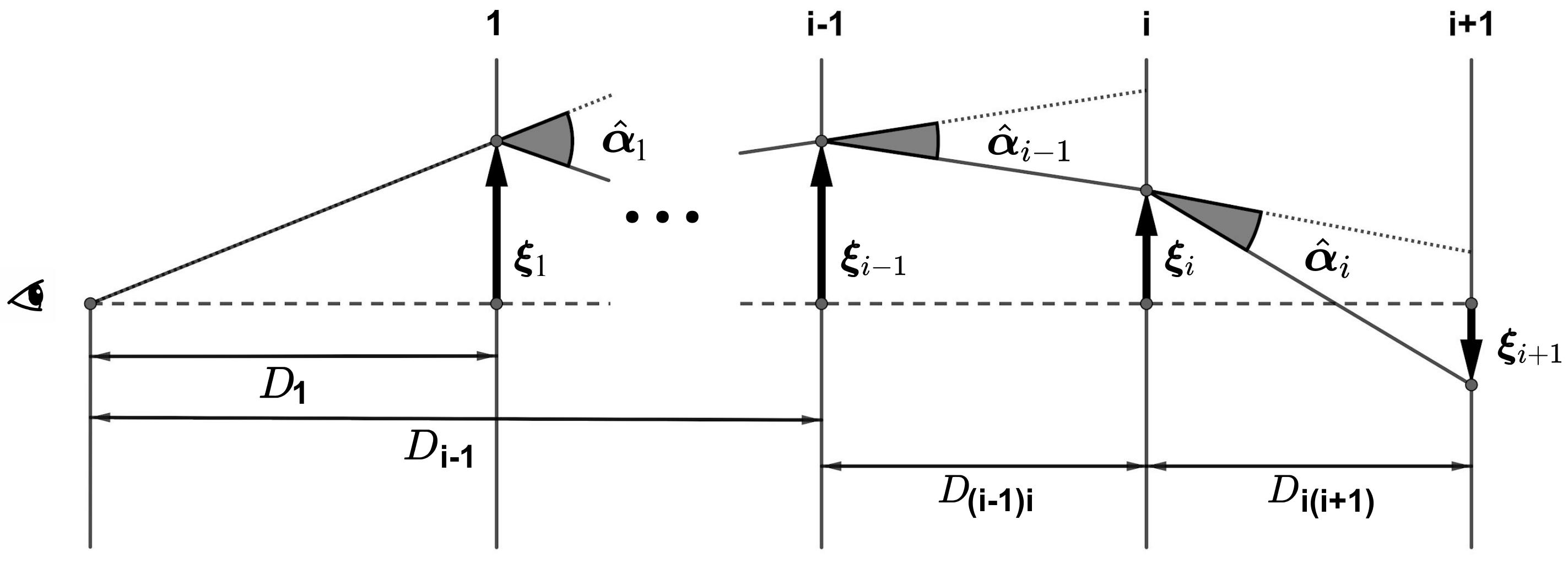}
    \caption{A schematic view of the multi-plane formalism, as described in Section~\ref{sec:theory-mpl}. A light ray (solid black line) experiences a deflection only when it passes through a lens plane (vertical solid grey lines). The deflection angle $\hat{\alpha}_{i}$ is the actual deflection of a ray passing through the $i$th lens plane, calculated from the surface density $\Sigma_{i}$ on the $i$th lens plane. Using the deflection angle $\hat{\alpha}_{i}$ and the position of the intersection of the light ray at the $(i-1)$th lens plane, $\bm{\xi}_{i-1}$, and that at the $i$th lens plane, $\bm{\xi}_i$, the physical position of the intersection at the $(i+1)$th plane, $\bm{\xi}_{i+1}$, can be obtained.}
    \label{fig:multiple_lens_planes}
\end{figure*}

In the case of multiple lens planes, the lens equation must be modified to account for multiple deflections;
\begin{equation}
 \bm{\beta} = \bm{\theta}-\sum^{N}_{i=1} \bm{\alpha}_{i}(\bm{\theta}_{i})\,\,,
 \label{eq:lensing_eq_mp}
\end{equation}
where the quantities retain their definition from the single lens plane case but now take on a subscript referring to a specific lens plane.
We consider N mass distributions, each characterised by a surface mass density $\Sigma_i$, at redshift $z_i$, ordered such that $z_i < z_j$ for $i < j$ and such that the source has a redshift $z_s > z_N$. The physical distance, $\bm{\xi}_j$, of the intersections on the lens planes from the optic axis (i.e., the impact parameters) are then
\begin{equation}
    \label{eq:xi_j}
    \bm{\xi}_j = \frac{D_j}{D_1} \bm{\xi}_1 - \sum^{j-1}_{i=1} D_{ij} \hat{\bm{\alpha}}_i (\bm{\xi}_i)\,\,,
\end{equation}
where $D_{i}$ is the angular diameter distance from the observer to each lens plane,  $D_{ij}$ (such that $i<j$) is the angular diameter distance from the $i$th lens plane to the $j$th lens plane and $\hat{\bm{\alpha}}_i$ is the deflection angle at the $i$th lens plane (see Fig.~\ref{fig:multiple_lens_planes}). For simplicity, we convert the physical distance to angular positions on the sky $\bm{\theta}_i = \bm{\xi}_i/D_i$ and the deflection angles to effective movements on the sky
\begin{equation}
    \label{eq:ai-ah}
    \bm{\alpha}_i = \frac{D_{is}}{D_s}\hat{\bm{\alpha}_i}\,\, ,
\end{equation}
where $D_{is}$ is the angular diameter distance from the $i$th lens plane to the source plane. By defining a factor $B_{ij}$
\begin{equation}
    \label{eq:gamma-ij}
    B_{ij} = \frac{D_{ij}D_{s}}{D_{j}D_{is}}\, ,
\end{equation}
eq.~\ref{eq:xi_j} becomes
\begin{equation}
    \label{eq:xi_j-new}
    \bm{\theta}_j = \bm{\theta}_1 - \sum^{j-1}_{i=1} B_{ij} \bm{\alpha}_i (\bm{\theta}_i)\, .
\end{equation}
In particular, for $j = N + 1 = s$, $B_{is} = 1$, thus,
\begin{equation}
    \label{eq:lq-mpl-beta}
    \bm{\beta} \equiv \bm{\theta}_{N+1} = \bm{\theta}_1 - \sum^{N}_{i=1} \bm{\alpha}_{i}(\bm{\theta}_i)\, .
\end{equation}

The delay of the arrival time of a deflected light path compared to a straight light path is the integral of the time difference along the line-of-sight though all lens planes. For instance, the time delay created by lens plane $i$ and $j$ is
\begin{equation}
    \label{eq:tau-ij-mpl}
    \tau_{ij} (\bm{\theta}_i, \bm{\theta}_j) = \frac{1 + z_i}{c}\frac{D_i D_j}{D_{ij}} \left[\frac{1}{2} (\bm{\theta}_i - \bm{\theta}_j)^2 - B_{ij} \psi(\bm{\theta}_i) \right]\, ,
\end{equation}
where the first term is the geometric delay and the second is the gravitational delay. Replacing $j$ with $i+1$ and summing over all time delays gives the total time delay through the whole line-of-sight,
\begin{equation}
    \label{eq:tau-all-mpl}
    \tau(\bm{\theta}_1,...,\bm{\theta}_N,\bm{\beta}) = \sum^{N}_{i=1} \tau_{i,i+1}(\bm{\theta}_i, \bm{\theta}_{i+1})\, .
\end{equation}
Therefore, similar to the case of a single lens plane, the time delay between two separate lensed images A and B can be given by
\begin{align}
    t_{\rm AB} &\equiv\tau_{\rm A} - \tau_{\rm B} \nonumber \\
    &= \sum^{N}_{i=1} \tau_{i,i+1}(\bm{\theta}_{{\rm A}, i}, \bm{\theta}_{{\rm A}, i+1}) - \sum^{N}_{i=1} \tau_{i,i+1}(\bm{\theta}_{{\rm B}, i}, \bm{\theta}_{{\rm B}, i+1})\,\,,
\end{align}
which means that deflection fields, lensing potentials and the angular positions of the intersections on the lens planes are all required for the calculation of time delays in multiple lens plane systems. In section 3, we discuss how we construct a light cone and model the lenses to obtain the information required to implement time-delay simulations with multiple lens planes.



\section{Simulations}
\label{sec:sims}
To quantify the influence of galaxies along the line-of-sight on measuring $H_0$ with strong lensing time-delays, we generated simulated images following the formalism in Sec.~\ref{sec:theory} for both single and multiple lens planes with a strong lensing simulation pipeline named \textsc{PICS} \citep{Li2016}. In this section, we describe the simulations used and how the lens equations are solved using a triangle-mapping algorithm.

\subsection{Semi-Analytic Lightcones}
\label{sec_lightcones}

For creating light cones with realistic spatial and redshift distributions of the galaxies, we extract light cones from the CosmoDC2 synthetic source catalogue \citep{CosmoDC2-2019}. Designed for an LSST data challenge project, it is established upon a large cosmological simulation called The Outer Rim Simulation run by the Argonne Cosmology Group using the Hybrid/Hardware Accelerated Cosmology Code \citep[HACC,][]{Habib2016}. CosmoDC2 covers 500 square degrees in the redshift range $0.0 \leq z \leq 3.0$ and is complete to a magnitude depth of 28 in the r-band. Each
galaxy is characterised by a multitude of properties including stellar mass, morphology, spectral energy distributions, broadband filter magnitudes, host halo information and weak lensing shear.

The light cones for each of our strong lensing simulations are cut out from the full light cone of CosmoDC2. Each extracted light cone is centred on a bright central galaxy (BCG) identified in the cosmoDC2 catalogue since these massive central elliptical galaxies are likely strong lensing candidates. Each BCG forms the primary lens mass in its corresponding light cone (see Section \ref{sec:ray-tracing}). The field of view of the light cones is $20''\times20''$, and the corresponding simulated images are $512\times512$ pixels in size. To focus on the impact of line-of-sight galaxies, we select light cones with the primary lens located in the redshift range $z_d = 0.5\pm0.01$ and we assume a fixed source redshift of $z_s = 2.0$. We calculate the Einstein radius of the primary lens of each light cone and then discard light cones that yield Einstein radii outside the range of $[1.3'', 2.4'']$. The lower limit avoids resolution issues encountered by ground-based telescopes/surveys (such as CFHT, DES, and LSST) and the upper limit discards systems which give year-like time delays. In total, we selected 500 light cones adhering to these criteria (although this is ultimately reduced further by additional selection criteria - see the following section and Section \ref{sec:results}). Furthermore, within each light cone, we remove any deflectors with Einstein radii larger than $0.3''$ to concentrate our study on the effects of secondary perturbations to the lensing potential. The substructures of the primary lens are also not included so that our analysis solely concentrates on the influence of line-of-sight structures.

\begin{figure*}
    \centering
    \includegraphics[width=0.75\textwidth]{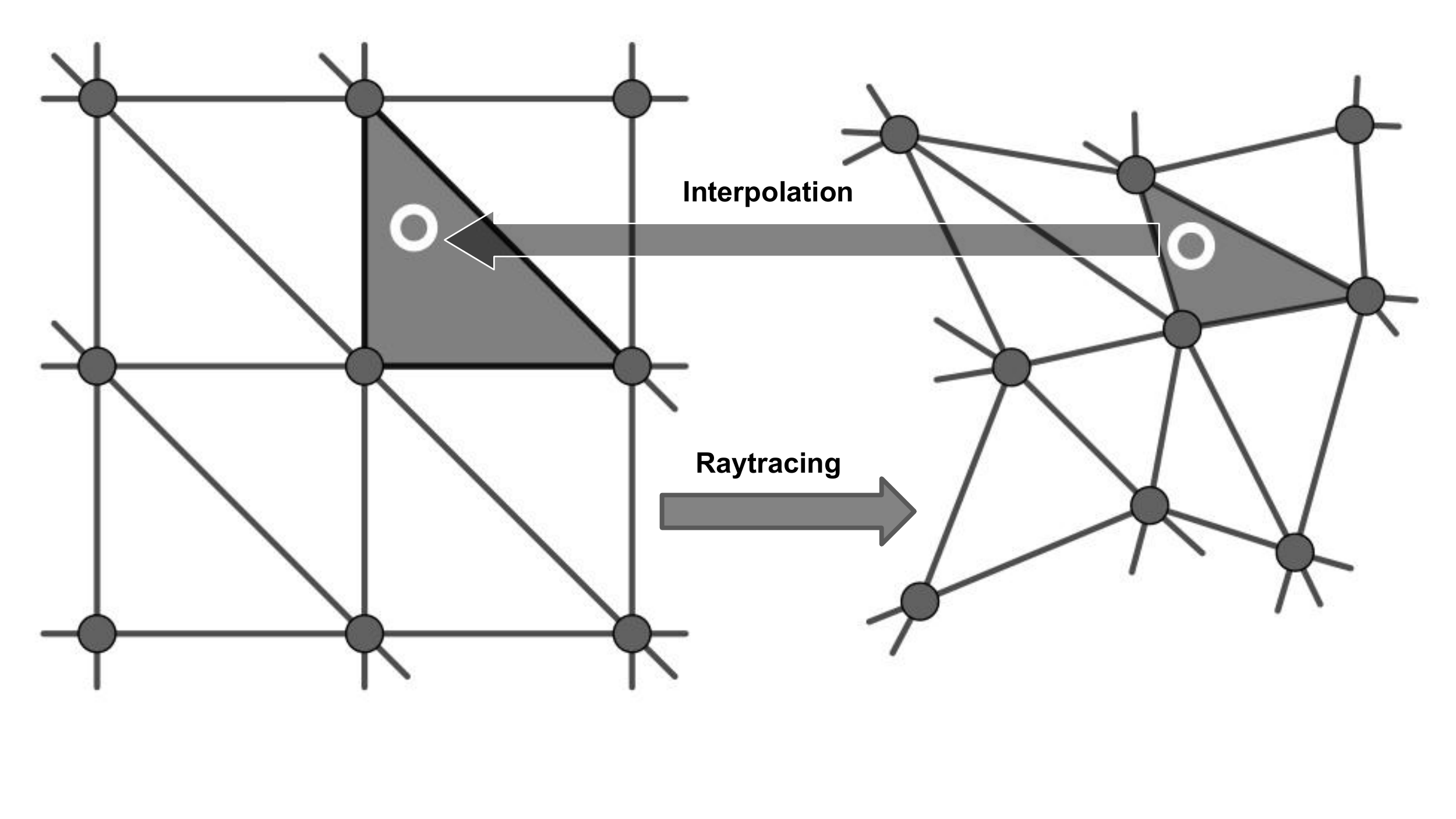}
    \vspace{-5mm}
    \caption{The Interpolation scheme used for determining image positions of point sources. The regular grid of rays in the image plane (left filled circles) is used to partition the image plane into triangles (grey lines in the left panel). The image positions (the open white circle in the left panel) of a source inside a triangle (the grey triangle in the right panel) formed by the backtraced rays on the source plane (grey filled circles in the right panel) is then determined by using linear interpolation in the barycentric coordinates.}
    \label{fig:root_finding}
\end{figure*}

\subsection{Ray-tracing Simulations}
\label{sec:ray-tracing}

For each light cone, we run two sets of simulations for generating the lens time delays. The first set includes only a single lens plane containing the primary lens galaxy. In this set, the omitted line-of-sight halos are approximated with a constant external convergence, $\kappa_{\rm ext}$, and a constant external shear, $\gamma_{\rm ext}$, in the lens model when computing deflection angles. For each light cone, we estimate the value of $\kappa_{\rm ext}$ and $\gamma_{\rm ext}$ by tracing multiple rays throughout it as described in more detail below. In the second set of simulations, we include all halos in the light cone and use a separate lens plane for each halo including the primary lens.

In both sets of simulations, we assume a singular isothermal ellipsoid (SIE) density profile for all halos (although in our lens modelling, we use a more general elliptical power-law profile; see Section \ref{sec:lens-modelling}). The SIE profile, which provides a realistic model for the total mass profile of real elliptical galaxies \citep{Koopmans2006,Bolton2012,Shu2016b}, has deflection angles given by \cite{Kormann1994,Keeton2001},
\begin{equation}
    \label{eq:sie-lq-x1}
    \alpha_x \equiv \psi_x = \frac{bq}{\sqrt{(1-q^2)}}\tan^{-1} \left[ \frac{\sqrt{1-q^2}\theta_x}{\phi}\right]\,\,,
\end{equation}

\begin{equation}
    \label{eq:sie-lq-x2}
    \alpha_y \equiv \psi_y = \frac{bq}{\sqrt{(1-q^2)}}\tanh^{-1} \left[ \frac{\sqrt{1-q^2}\theta_y}{\phi}\right]\,\,,
\end{equation}
where $\phi^2 = q^2 x^2 + y^2$, $q$ is the minor to major axis ratio and $b$ is an effective factor to represent Einstein radius,
\begin{equation}
    \label{eq:sie-re-0}
    b = \frac{4 \pi}{\sqrt{q}} \left( \frac{\sigma}{c}\right)^2 \frac{D_{ls}}{D_s}\,\,.
\end{equation}
In the case of circular lenses, $b$ can be calculated from the velocity dispersion. The lensing potential can be computed according to the relationship between the lensing potential and the deflection field of SIE model \citep{Keeton2001},
\begin{equation}
    \label{eq:sie-psi}
    \psi(\theta_x, \theta_y) = \theta_x \psi_x + \theta_y \psi_y \,\, .
\end{equation}

The complete parameter set required by equations~(\ref{eq:sie-lq-x1} $-$~\ref{eq:sie-psi})  is $\{x_1, x_2, \sigma_{v}, q, \Theta, z_d\}$, where $(x_1, x_2)$ is the angular position of the SIE profile centre with respect to the centre of the field of view, $\sigma_{v}$ is the velocity dispersion of the lens, $q$ is the ellipse axis ratio, $\Theta$ is the position angle of the ellipsoid and $z_d$ is the redshift of the deflector. The parameters ${x_1, x_2, q, \Theta, z_d}$ are taken directly from the cosmoDC2 catalogue.
$\sigma_v$ is derived from the $L-\sigma$ scaling relation from the bright sample of \cite{Parker2007} given by
\begin{equation}
\sigma_v = 142 \left( \frac{L}{L_{\star}} \right)^{(1/3)}~{\rm km~s}^{-1}\,\, ,
\end{equation}
where, $\log_{10}(L/L_{\star}) = -0.4(mag_r - mag_{r \star})$, and $mag_r$ is the apparent $r$-band magnitude of the galaxy given by the cosmoDC2 catalogue. We adopt the assumption in \cite{SpacewarpsII-2016} that $mag_{r \star}$ evolves with redshift as $mag_{r \star} = +1.5(z-0.1)-20.44$ \citep{Faber2007}.

Sources are described by the parameter set $\{y_1, y_2, m_s, z_s\}$, where $(y_1, y_2)$ is the angular position of the source with respect to the optic axis, $m_s$ is the apparent $r$-band magnitude of the source and $z_s$ is the redshift, fixed to $z_s = 2$. The angular positions are randomly sampled in the source plane in the vicinity of the caustic structures. We only retain simulated data in which quadruply-lensed images are produced in both versions of a given light cone, i.e. both the single and the multiple lens-plane version. This reduces our initial selection of 500 light cones (see Section \ref{sec_lightcones}) to 400.

With a fully parametrically-defined light cone, the simulated lensed images can be produced by ray-tracing and image-finding. For our single lens-plane simulations, we determine $\kappa_{\rm ext}$ and $\gamma_{\rm ext}$ in the following manner. First, we trace rays through a given light cone from the image plane, computing the deflections caused by all halos (including the primary lens), each in their own lens plane. To obtain $\gamma_{\rm ext}$, along each ray, we compute the cumulative external shear from all halos. We take $\gamma_{\rm ext}$ to be the median of the distribution of values of the cumulative external shear along different rays in the light cone. For the external convergence, along each ray, we compute an 'external halo convergence' by summing $\kappa$ as given by Eq.~\ref{eq:kappa} for all secondary halos excluding the primary lens halo. This external halo convergence ignores the divergence caused by voids and so we must apply a correction to obtain $\kappa_{\rm ext}$. The correction uses the results of \cite{Collett2013} who showed that $\kappa_{\rm ext}$ can be obtained by subtracting the median convergence along random sight lines from the external halo convergence. The resulting $\kappa_{\rm ext}$ has an uncertainty associated with it due to the scatter in the relationship between the two quantities, but negligible bias. Firing rays along random lines-of-sight in our light cones and computing the convergence, again using Eq.~\ref{eq:kappa}, yields a value of $\kappa_{\rm corr}=0.048$. When correcting the external halo convergence, we distribute $\kappa_{\rm corr}$ across all lens planes according to the lensing weights ($D_{ds}D_{d}/D_{s}$) for each plane and subtract them separately.

\begin{figure}
    \centering
    \includegraphics[width=0.47\textwidth]{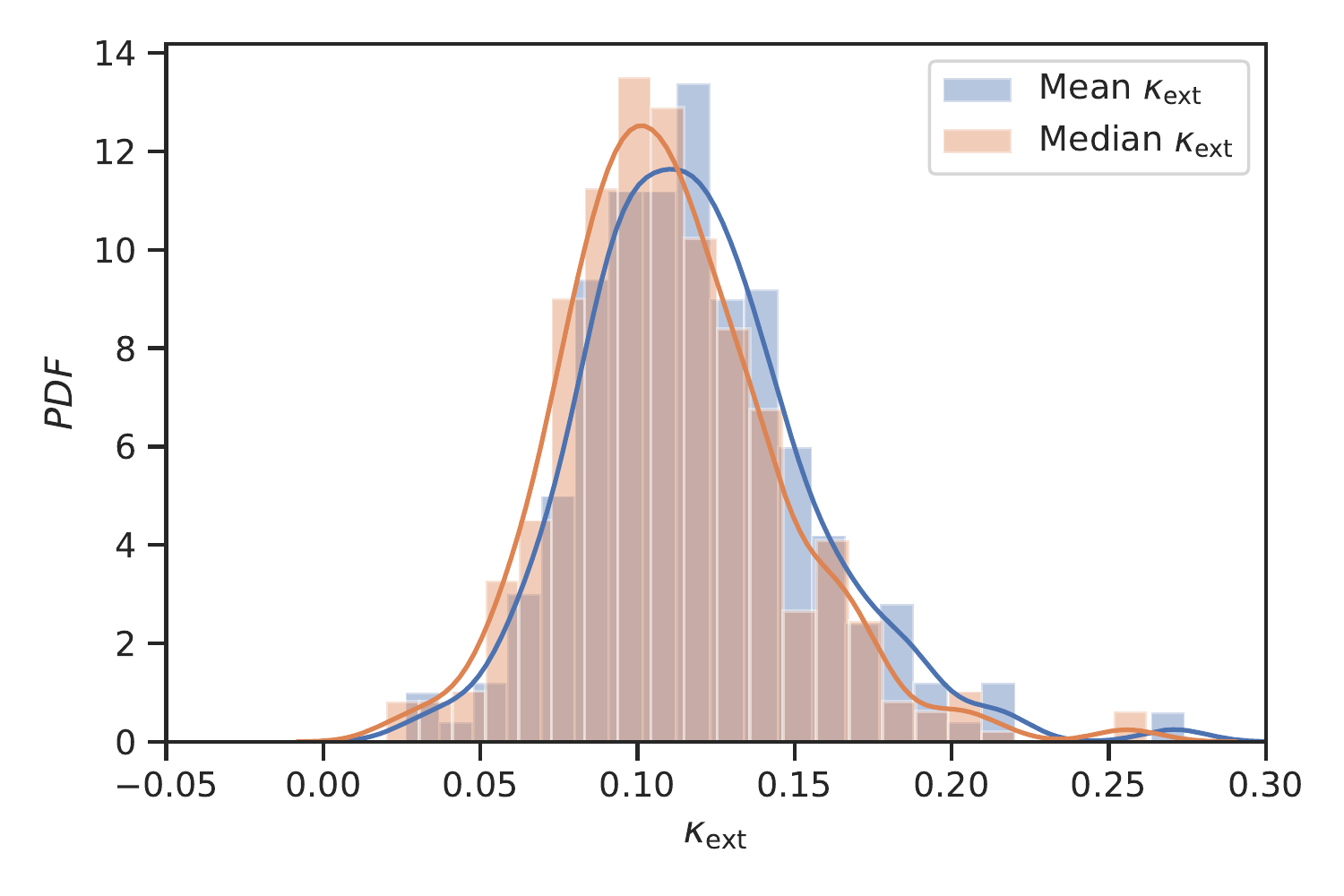}
    \caption{The distribution of the mean (blue) and median (orange) convergence of all fully ray-traced light cones used in this work. The blue and orange curves show a smoothed version of the distributions calculated using kernel density estimates.}
    \label{fig:extKappa}
\end{figure}

Figure \ref{fig:extKappa} shows the probability distribution functions (PDFs) of the mean and median values of $\kappa_{\rm ext}$ across all light cones obtained in the manner described. We note that our peak of $\kappa_{\rm ext} \simeq 0.1$ is higher than that of previous studies, for example, peaks of $0.075$ and $0.05$ in \cite{Suyu2013} and \cite{McCully2017} respectively. We attribute this mainly to our selection of BCGs from cosmoDC2 and their location within more over-dense galaxy groups. Secondary effects also likely include a difference in mass models and simulated light cones. Nevertheless, many of our light cones yield external convergences that are consistent with these studies and so in our analysis, we explore how  inferred values of $H_0$ vary with varying $\kappa_{\rm ext}$.

With $\kappa_{\rm ext}$ determined, we include it in the primary lens model for the single-plane simulations and calculate maps of the deflection angle and the lensing potential. The lensing equation in Eq.~\ref{eq:lensing_eq_slp} is used to map the image plane back to the source plane. Since the sources in this paper are point sources, we have to adopt a triangle-mapping algorithm to solve the lensing equation. This is described further in Section~\ref{sec:image-finding}.

For the case of multiple lens-planes, we ray-trace through the whole light cone in the same manner as outlined above when computing the external halo convergence, placing each halo on its own lens plane. As Eq.~\ref{eq:tau-ij-mpl} shows, to calculate the total time delay, the deflection map and lensing potential for every lens plane must be computed. The intersections of the light rays traced from the image plane (given by Eq.~\ref{eq:xi_j-new}) are required for the calculation of the time delay between two lens planes. These are summed over all neighbouring pairs of lens planes to obtain the total time delay according to Eq.~\ref{eq:tau-all-mpl}. Again, for our adopted point source, we have to apply triangle mapping and barycentric interpolation to obtain the position of lensed images for a given source position on the source plane (see Section~\ref{sec:image-finding}). The same image-finding process is applied to locate the intersections of the light rays between neighbouring lens planes (see Eq.~\ref{eq:tau-ij-mpl}).

Since we are concerned purely with the effects of line-of-sight structure in this study, we have not included the effects of measurement error in our simulated data, i.e. we do not scatter any of the time delays, image positions or flux ratios. However, we do use priors in the modelling to allow exploration of parameter degeneracies. More details are given in Section~\ref{sec:lens-modelling}.

\subsection{Image Finding}
\label{sec:image-finding}
Since we are concerned with multiply-imaged point-like sources, e.g. AGNs or SNe, in this work, solving the lensing equation for point sources is a critical issue in the simulation. To determine the apparent positions of our point-sources, we make use of a triangle mapping technique described in \citep{Schneider1992}. First, a set of Delaunay triangles is constructed from a regular grid of image plane positions which define the intersections of light rays from the source (see Fig.~\ref{fig:root_finding}). These image plane vertices are then mapped to the source plane. Any image plane triangles which map to a triangle in the source plane containing the source position are identified. For each of these identified image-plane triangles, we compute the barycentric coordinate of the source position inside the corresponding source-plane-mapped triangle using the relation
\begin{equation}
\begin{pmatrix}
x_1 & x_2 & x_3\\
y_1 & y_2 & y_3\\
  1 &   1 &   1
\end{pmatrix}
\begin{pmatrix}
\lambda_1\\
\lambda_2\\
\lambda_3
\end{pmatrix}
=
\begin{pmatrix}
x_P\\
y_P\\
1
\end{pmatrix}
\end{equation}
where, $(x_P, y_P)$ are the Cartesian coordinates of the point source inside its triangle of vertices $(x_1, y_1)$, $(x_2, y_2)$, and $(x_3, y_3)$; the corresponding barycentric coordinates are $(\lambda_1, \lambda_2, \lambda_3)$. We then assume that the barycentric co-ordinates are conserved between the image and source planes and use them, with the vertices of the image-plane triangle to determine the position of each image of the source.

For the case of multiple lens planes, the intersections between the light rays from the source and the lens planes are required for the calculation of total time-delays. Hence, we need to ascertain all the intersections. If there are $N$ lens planes plus one source plane in the lensing system, there are $N$ parent triangles for the triangle on the source plane. Also, we assume the barycentric coordinates of the source are conserved in the source triangle and all parent triangles. Then the intersections can be obtained. The intersections on the first lens plane ($0$th plane in Fig.~\ref{fig:multiple_lens_planes}) are the positions of the lensed images.
\label{sims}

\section{Strong Lens Modelling}
\label{sec:lens-modelling}

We use the multi-purpose open-source lensing package
 \textsc{lenstronomy}\footnote{\url{https://github.com/sibirrer/lenstronomy}}
\citep{Birrer:2015rpa,Birrer:2018xgm} to measure $H_0$ from our simulated data.
For our lens modelling, instead of the SIE profile used to create our simulated data, we use the more general Singular Elliptical Power Law (SEPL) profile. The parameters of the SEPL are the Einstein radius, $\theta_{\rm E}$, the two components of complex ellipticity, $e_1$ and $e_2$, the SEPL power-law index, $\gamma$ and the co-ordinates of the SEPL centre, $(\theta_1,\theta_2)$. Also included as free parameters in the modelling are the co-ordinates of the source, $(\beta_1,\beta_2)$, in the source plane. Finally, we apply the SEPL model both with and without external shear (see below). We use the complex shear parameterised by $\gamma_{\rm ext,1}$ and $\gamma_{\rm ext,2}$. We apply generous uniform priors to all model parameters in \textsc{lenstronomy} as detailed in Table~\ref{table:prior}.

We model all four different combinations arising from the two lens model configurations (i.e., the SEPL with and without external shear) and the two sets of simulated data (i.e., the single and multiple lens plane light cones). We designate the simulations with a single lens plane as 'SGK' (SIE + $\gamma_{\rm ext}$ + $\kappa_{\rm ext}$) and those with the multiple lens plane as 'SL' (SIE + Lens planes). Similarly, we designate the lensing model that includes external shear as 'SG' (SEPL + $\gamma_{\rm ext}$) and that without as 'S'. The four combinations, labelling the simulation type first, are therefore referred to hereafter as 'SGK|S', 'SGK|SG', 'SL|S' and 'SL|SG'. Note that in all cases we fix $\kappa_{\rm ext}=0$ and retrospectively apply the correction to $H_0$ for external convergence determined from the simulated light cones following the procedure used by existing studies (see Section \ref{sec:results}). In cases where external shear is not included as a free parameter in the lens model (SGK|S and SL|S), we fix $\gamma_{\rm ext}=0$.

The simulated data that we fit with \textsc{lenstronomy} are the four image positions, the three flux ratios, and the three time delays. For optimisation of the lens model parameters and $H_0$, we use \textsc{lenstronomy}'s particle swarm optimiser (PSO) \citep{eberhart1995particle} since this technique performs well in lower dimensional parameter spaces such as ours \citep[see][]{2015ApJ...813..102B}. We apply the PSO with $200$ particles, a particle scatter of $1$, and a maximum number of iterations of $500$. These choices yield an acceptable computation time whilst still allowing a thorough exploration of the model parameter space.

\begin{table}
\centering
\begin{tabular}{p{9em} p{9.5em} p{5.5em}}
\toprule
\textbf{Model component}&\textbf{Parameter}&\textbf{Prior}\\
\midrule
Lens, Einstein radius    & $\theta_E$ (arcsec) & $\mathcal{U}(0.01, 10)$ \\
Lens, power index    & $\gamma$ & $\mathcal{U}(1.7, 2.3)$ \\
Lens, ellipticity    & $e_{1,2}$ & $\mathcal{U}(-0.5, 0.5)$ \\
Lens, position    & $\theta_{1,2}$ (arcsec) & $\mathcal{U}(-10, 10)$ \\
External shear & $\gamma_{\rm ext}$ & $\mathcal{U}(0.0, 0.5)$ \\
External shear angle & $\theta_{\gamma, {\rm ext}}$ (rad) & $\mathcal{U}(-\pi, \pi)$ \\
Source, position   & $\beta_{1,2}$ (arcsec) & $\mathcal{U}(-10,10)$ \\
Hubble-Lemaitre  constant        & $H_0$ (km/s/Mpc) & $\mathcal{U}(20, 120)$ \\
\bottomrule
\end{tabular}
\caption{Uniform priors applied to parameters in the lens modelling.}
\label{table:prior}
\end{table}

\section{Results}
\label{sec:results}

In carrying out the modelling, we find that not all measurements of $H_0$ obtained are valid. This is due to the limited precision of the simulations; when a source is almost coincident with the caustic in the source plane, the magnifications of the simulated lensed images become unreliable because of the finite image grid size, despite our interpolation. These problematic cases can be effectively removed by imposing a likelihood threshold of ${\rm log}(L) > -1000$. This further reduces our sample of 400 lens systems to 364, 372, 366, and 394 lenses in the cases of SGK|S, SGK|SG, SL|S, and SL|SG respectively. By applying this threshold in likelihood, we also remove poor fits arising from large perturbations from substructures not caught by the $0.3''$ cut in Einstein radius.

\begin{figure}
    \centering
    \includegraphics[width=0.47\textwidth]{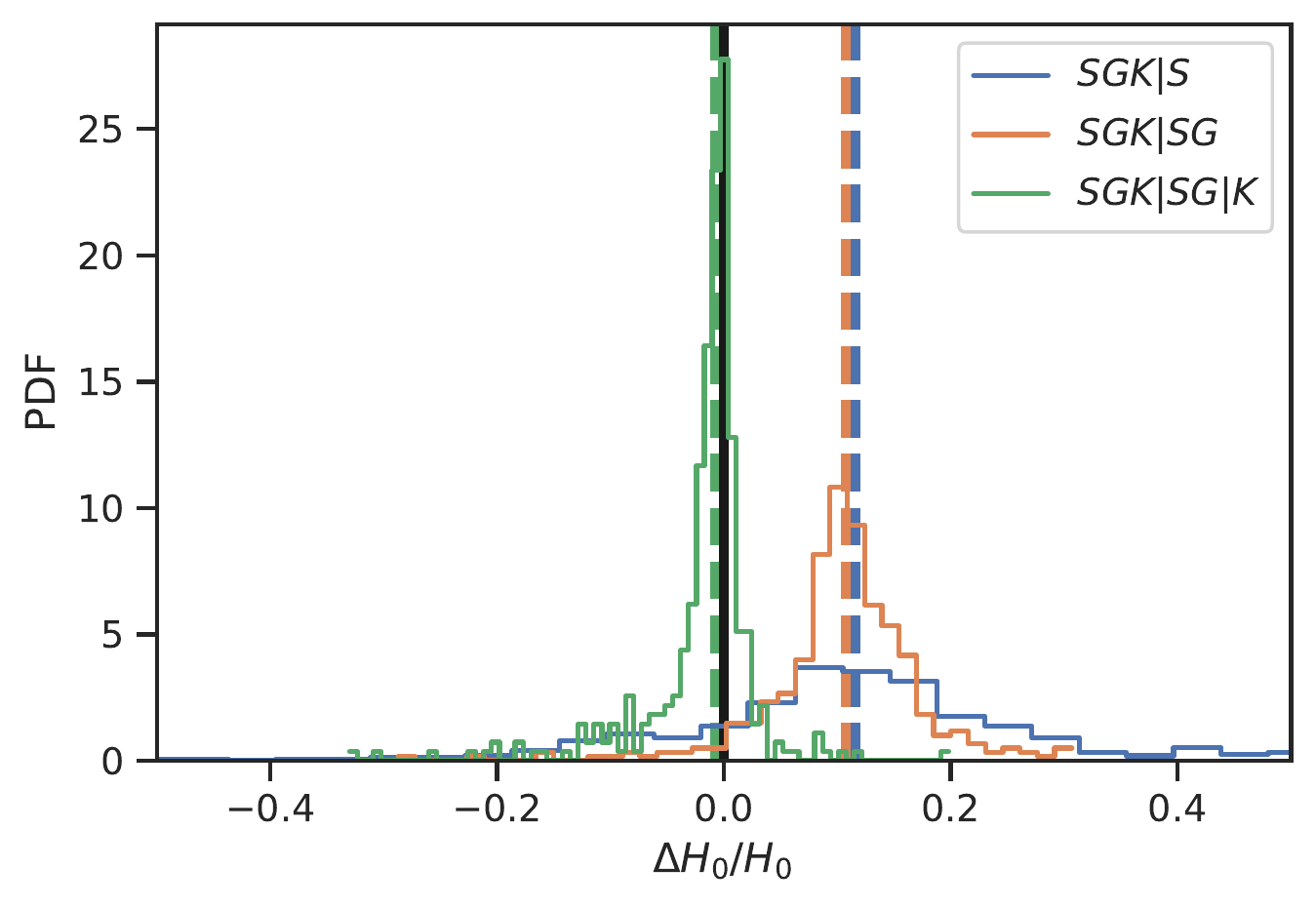}
    \caption{PDFs of the fractional differences between measured $H_0$ and the true value in the case of the simulations with constant $\kappa_{\rm ext}$ and $\gamma_{\rm ext}$. The blue histogram shows the PDF of fractional differences in $H_0$ with the single $SEPL$ mass model only. The orange histogram shows the PDF of fractional differences with the mass model of $SEPL$+$\gamma_{\rm ext}$, i.e. including external shear as a free parameter. The green histogram shows the corrected fractional differences of the orange histogram with constant $\kappa_{\rm ext}$ correction. The vertical dashed lines show the median of each PDF whilst the black vertical solid line is placed at zero bias.}
    \label{fig:dh0_PDF_kext}
\end{figure}

First we consider our analysis of the simulations created with LOS structure approximated by a constant external convergence and shear. Fig~\ref{fig:dh0_PDF_kext} shows the PDFs of the fractional difference between the input and inferred $H_0$ obtained for the two different lens models applied, i.e. the SEPL-only model (SGK|S) and the SEPL+$\gamma_{\rm ext}$ model (SGK|SG). Taking the median of each of these distributions, we find that without including any external convergence in the modelling, the inferred value of $H_0$ is biased high by $\sim 11$ per cent in both cases. The inclusion of external shear in the lens model reduces the spread of the distribution but does nothing to remove the bias.

Following the procedure commonly used in the literature to correct for external convergence effects \citep[see, for example][]{Suyu2017},
we apply a correction of $1 - \kappa_{\rm ext}$ (with $\kappa_{\rm ext}$ determined from the simulations as explained in Section \ref{sec:ray-tracing}) to the biased measurements of $H_0$ from the $SGK|SG$ configuration. The green histogram shown in Fig~\ref{fig:dh0_PDF_kext} shows the results of this correction. Clearly, the correction in this simplified case works well, recovering a median value of $H_0$ that differs from the input value by only $-0.7$ per cent.

\begin{figure*}
    \centering
    \includegraphics[width=1.0\textwidth]{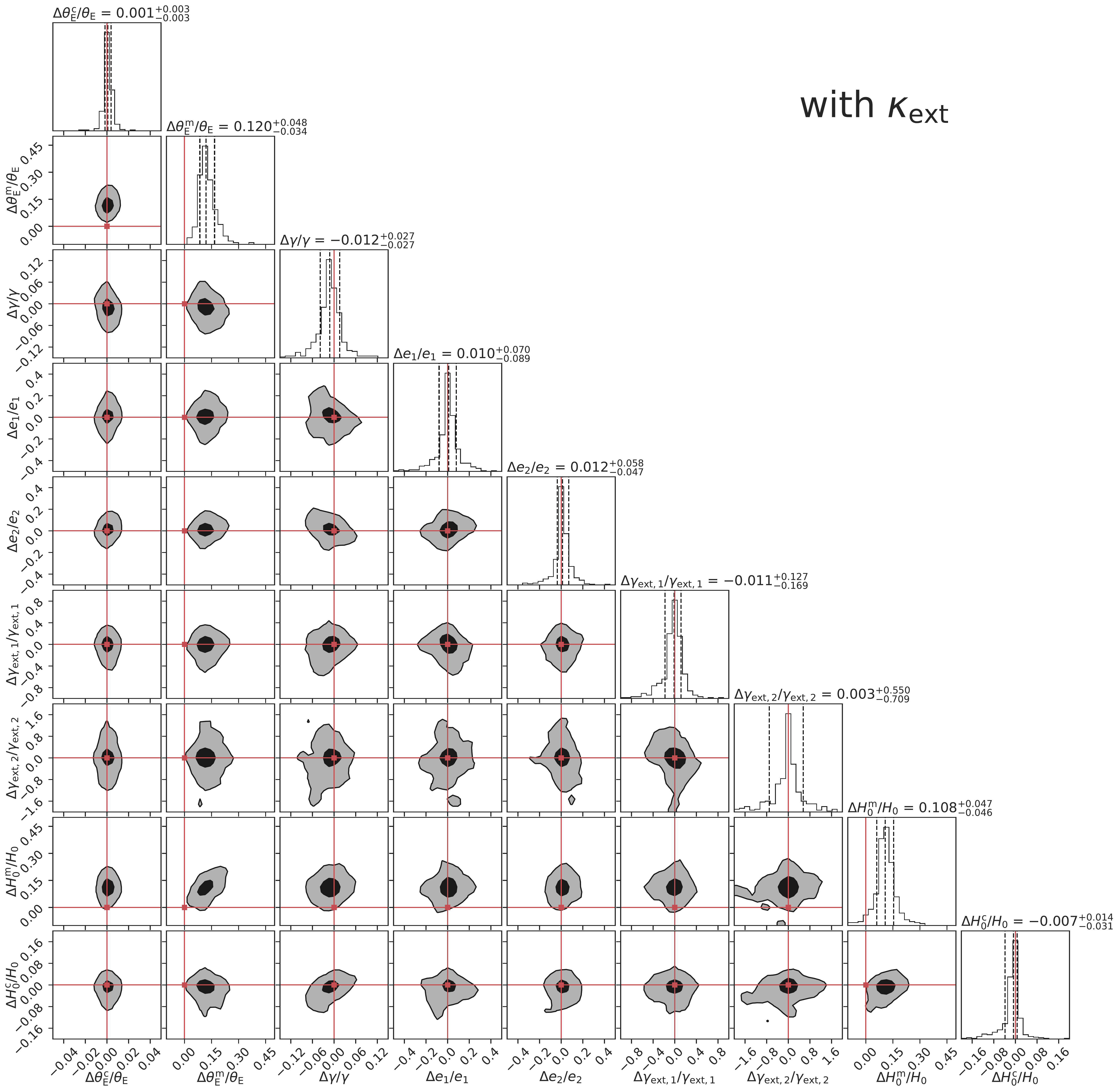}
    \caption{Corner plot showing the distribution of best-fit parameters of all 372 systems simulated with a single lens plane and uniform external convergence and shear. The plot includes the Einstein radius and $H_0$ corrected by the simplistic factor of $1-\kappa_{\rm ext}$. These are denoted $\theta_{\rm E}^{\rm c}$ and $H_{0}^{\rm c}$ respectively. $\gamma$ is the power index of the $SEPL$ mass model, $e_1$ and $e_2$ are the two components of the complex ellipticity of lenses, $\gamma_{\rm ext,1}$ and $\gamma_{\rm ext, 2}$ are the two components of the complex external shear, $H_0^{\rm m}$ is the best-fit uncorrected Hubble constant and $H_0$ is the input Hubble constant. The contours show the 1- and 2-sigma confidence intervals. The plot is created with \textit{Corner.py} \citep{corner}.}
    \label{fig:hExtKappa}
\end{figure*}

In Fig~\ref{fig:hExtKappa}, we show the two-dimensional probability distributions of all parameter pair combinations for the $SGK|SG$ configuration. Note that in addition to the bias in $H_0$, there is also a similar bias in the inferred Einstein radius, $\theta_{\rm E}$. This is a result of the strong degeneracy between $\theta_{\rm E}$ and $H_0$ caused by the fact that the external convergence impacts both quantities by the same factor of  $1-\kappa_{\rm ext}$. As Fig~\ref{fig:hExtKappa} shows, correcting $\theta_{\rm E}$ by the factor $1-\kappa_{\rm ext}$ (to give the quantity $\theta^{\rm c}_{\rm E}$ in the figure), the input value of the Einstein radius is reliably recovered.

\begin{figure}
    \centering
    \includegraphics[width=0.47\textwidth]{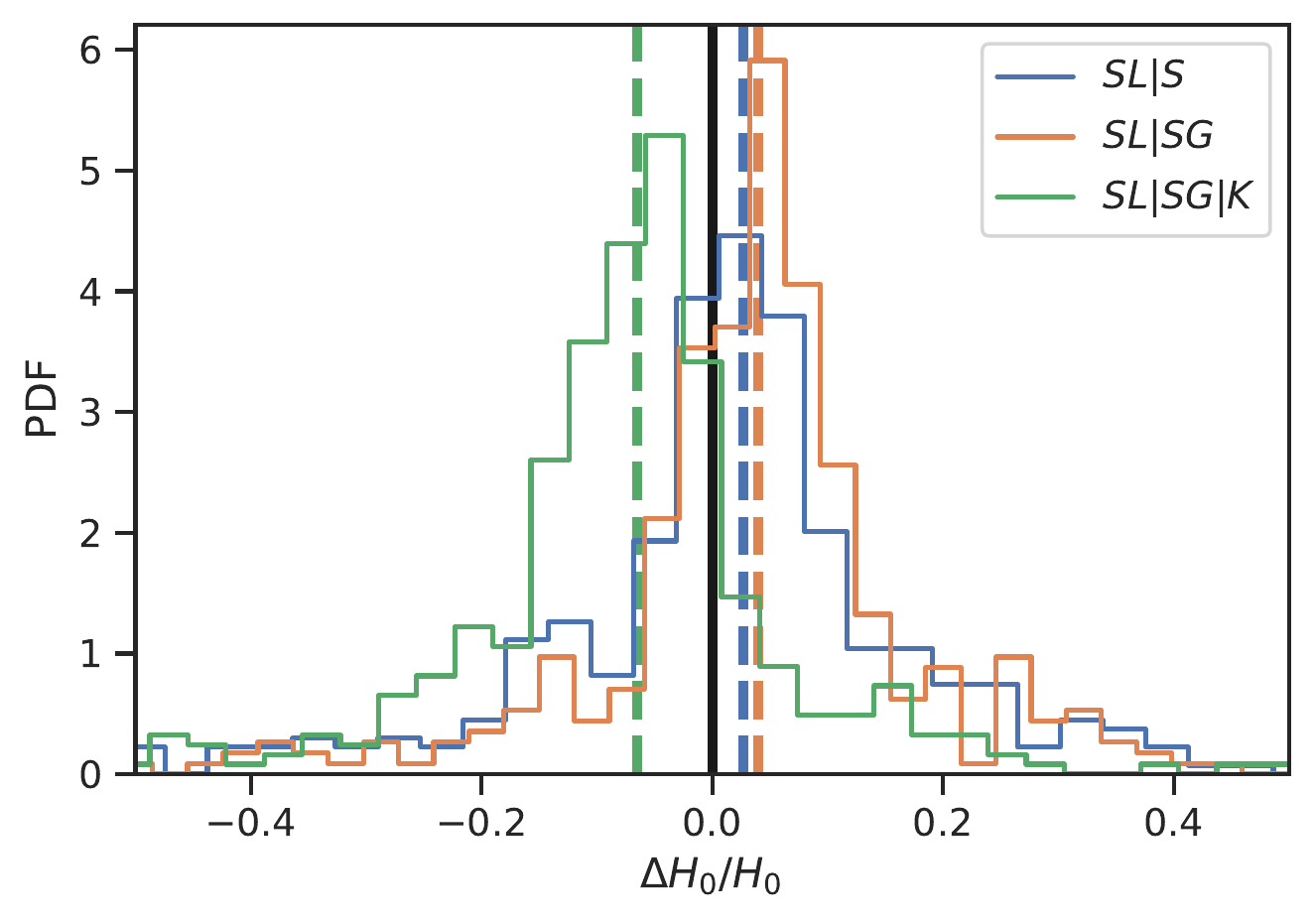}
    \caption{The same as Fig~\ref{fig:dh0_PDF_kext}, except using the fully ray-traced simulations containing line-of-sight halos.}
    \label{fig:dh0_PDF_SL}
\end{figure}

Second, we consider our modelling of the simulations created with the full light cones containing halos (i.e., the cases of $SL|S$ and $SL|SG$). Fig.~\ref{fig:dh0_PDF_SL} shows the distribution of inferred values of $H_0$ for both cases. This time, we find that the biases in inferred $H_0$ are significantly smaller than the biases observed with the single lens plane light cones. Now, we find a median value that is 3 and 4 percent higher than the input value of $H_0$ for the $SL|S$ and $SL|SG$ cases respectively. Once again, the inclusion of external shear in the lens model does little to improve the bias. Furthermore, the inclusion of external shear does not reduce the scatter in inferred values of $H_0$, unlike the single lens plane modelling. Fig.~\ref{fig:dh0_PDF_SL} also shows the histogram of inferred $H_0$ from the modelling that includes external shear (SL|SG) corrected by $1-\kappa_{\rm ext}$, where again, $\kappa_{\rm ext}$ is determined from ray tracing through the light cone. This time, the correction is too severe and leads to an underestimation of $H_0$ such that the corrected distribution has a median that is offset by -7 per cent from the input value. We therefore conclude that statistically, the $1 - \kappa_{\rm ext}$ correction can not be reliably used to account for clumpy external convergence.

\begin{figure*}
    \centering
    \includegraphics[width=1.0\textwidth]{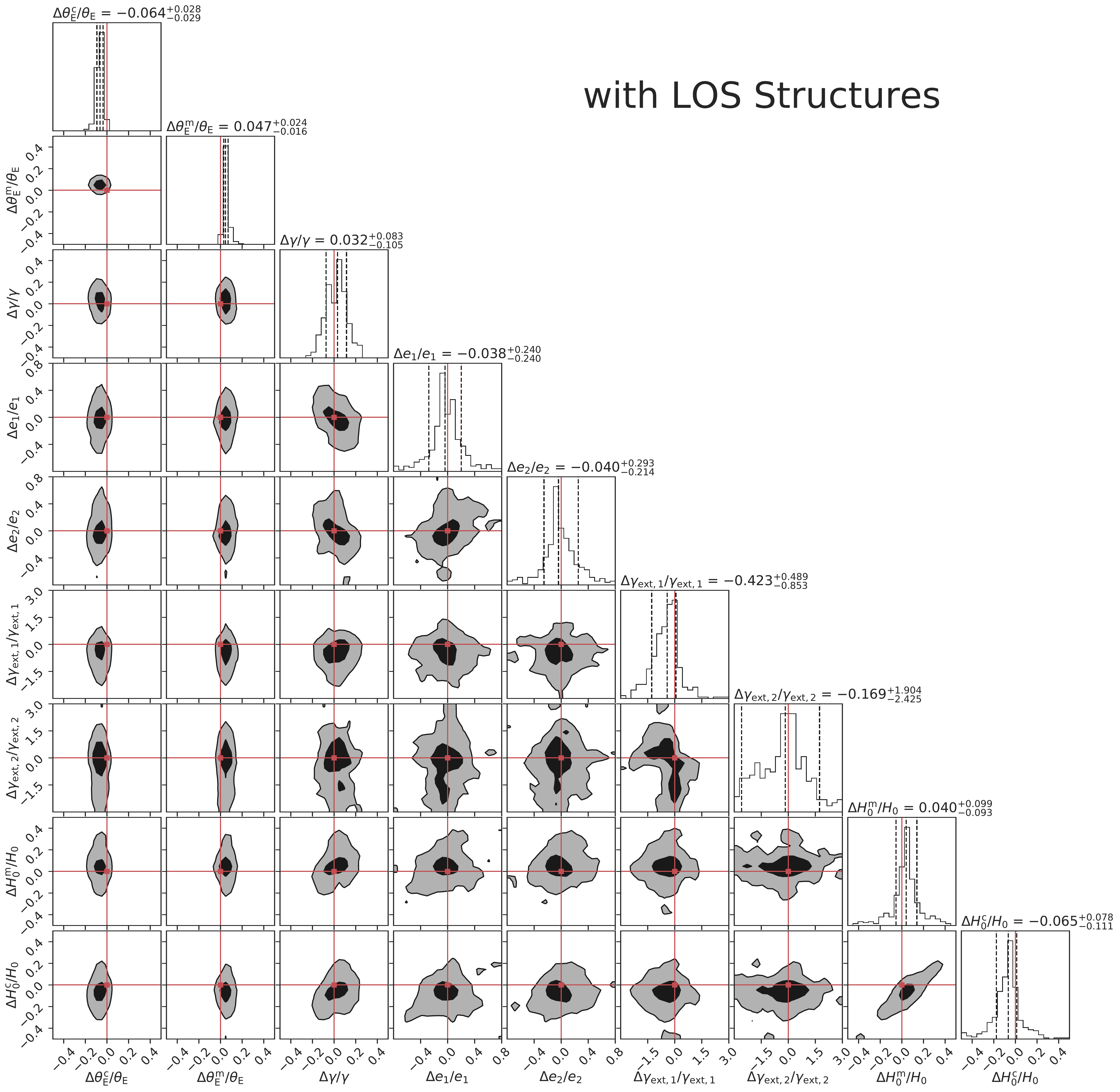}
    \caption{Corner plot showing the distribution of best-fit parameters of all 394 systems simulated by ray tracing through light cones containing line of sight halos. All parameters are the same as those in Fig~\ref{fig:hExtKappa} and the contours again show the 1- and 2-sigma confidence intervals.}
    \label{fig:hLosKappa}
\end{figure*}

Similar to Fig~\ref{fig:hExtKappa}, Fig~\ref{fig:hLosKappa} shows the two-dimensional probability distributions of all parameter pair combinations for the SL|SG configuration. Again, the figure includes both $H_0^{\rm c}$ and $\theta^{\rm c}_{\rm E}$, the inferred values of $H_0$ and Einstein radius corrected by $1-\kappa_{\rm ext}$. This time, however, the degeneracy between $H_0$ and $\theta_{\rm E}$ has been removed by the more complex lens geometry caused by the line-of-sight structure; clumpy external convergence affects the time delays in a different way to the way in which it affects the inferred Einstein radius, unlike when a uniform external convergence is assumed. In the same way that the inferred $H_0$ is not biased as high with the full light cones, neither is the inferred Einstein radius and so the correction provided by the factor of  $1-\kappa_{\rm ext}$ is also too severe and also results in a bias of -7 per cent from the input value on average.

\begin{figure}
    \centering
    \includegraphics[width=0.47\textwidth]{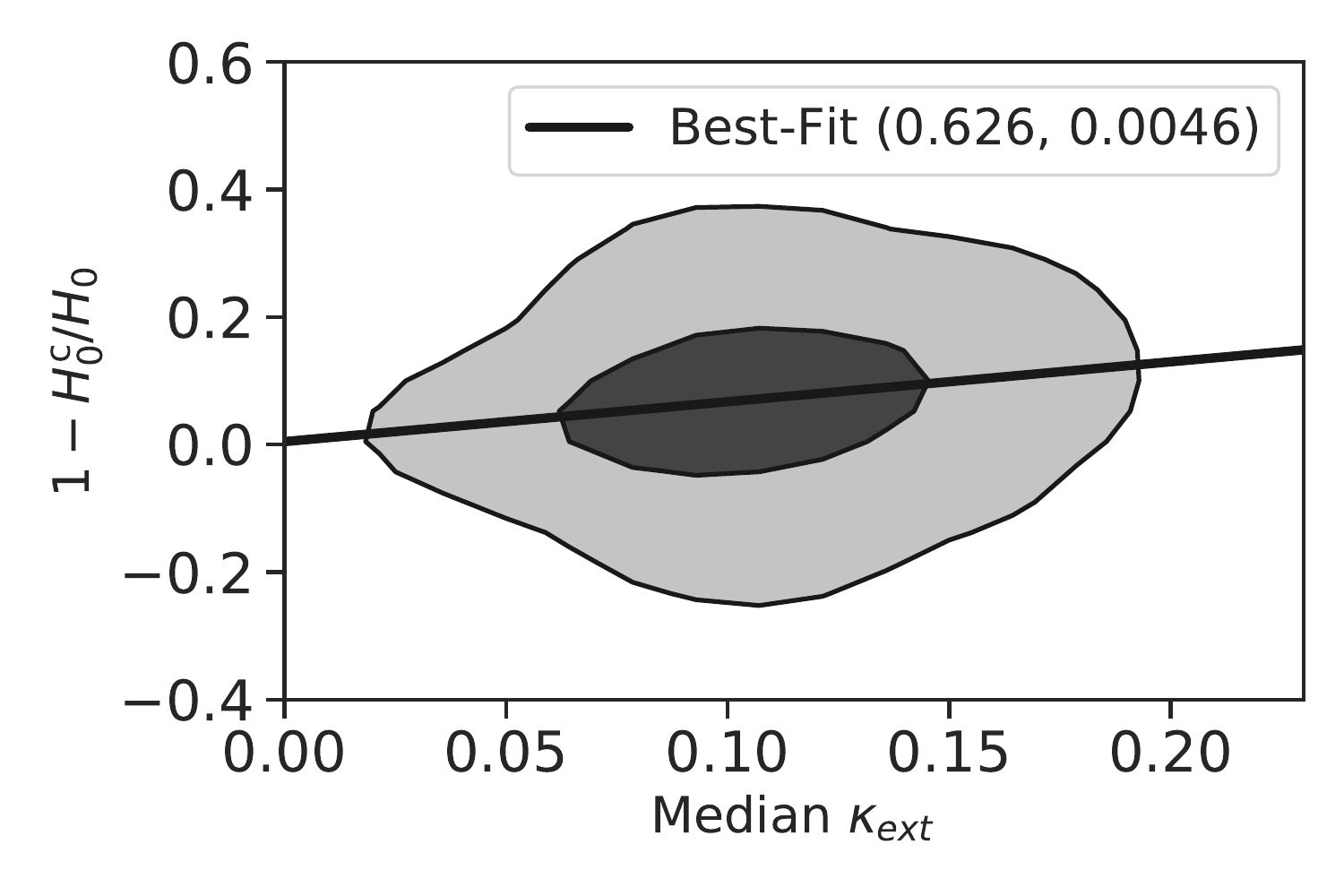}
    \caption{The relationship between the fractional bias seen in the corrected value of $H_0$, $H_0^{\rm c}$, and the median external convergence measured across all 394 fully ray-traced light cones containing line of sight halos. The contours show the 1- and 2-sigma confidence intervals and the black line shows the best-fit linear relationship which exhibits almost exact proportionality: $1- H_0^{\rm c} / H_0 = 0.626 \kappa_{\rm ext} - 0.005$.}
    \label{fig:biasVSKext}
\end{figure}

Since our simulations span a range of lens systems each with a different median external convergence, $\kappa_{\rm ext}$, we can investigate whether there is any correlation between the bias we see in inferred $H_0$ and $\kappa_{\rm ext}$. Identifying such a correlation might instruct future studies on how best to minimise the bias. Fig.~\ref{fig:biasVSKext} shows the scatter plot of the bias in inferred $H_0$ versus $\kappa_{\rm ext}$ for each lens system with the SL|SG configuration. As the figure shows, there is a positive correlation such that the fractional bias in $H_0$ due to the over-correction correlates with the median external convergence. The scatter plot can be fitted using a linear function $\Delta H_0 / H_0 = 0.626 \kappa_{\rm ext} - 0.005$. Unsurprisingly, selecting a lens system in an environment with a stronger level of external convergence is more likely to bias the value of $H_0$ inferred from that system.

\section{Discussion and Conclusions}
\label{sec:conclusion}

To quantify the influence of secondary deflectors on the measurement of $H_{0}$ with strong lensing time delays, we have simulated approximately 800 galaxy-scale strong lensing systems with quadruply-lensed variable point sources; half of these were created with a primary lens and line-of-sight halos and half with the same primary lens plus a constant external convergence and shear. The light cones were extracted from a semi-analytic model based on the Outer Rim large-scale cosmological simulation and are centred on the location of central galaxies of groups of galaxies. In the simulations constructed with external convergence and shear, we used a single lens-plane located at the redshift of the primary lens galaxy whereas in the simulations containing halos, each halo has its own lens plane. Using an SIE mass profile for the primary lens galaxy and the halos, and an interpolative mapping method to refine the location  of the lensed point source images, we generated time delay data. This time-delay data was then modelled using \textsc{lenstronomy} to estimate $H_0$ with a singular ellipsoidal power law lens profile and external shear and compared to the known input value of $H_0$.

Our main conclusion is that incorporating constant external convergence in the modelling only works reliably if the lensed time delays are subjected to a uniform external convergence. If time-delays are subjected to perturbations due to halos lying close to the line-of-sight as expected in the real Universe, and no correction for external convergence is made in the modelling, the inferred value of $H_0$ is over-estimated by approximately $4$ per cent on average. However, if a constant external convergence is incorporated in the lens model with a normalisation set by the median or mean convergence of the line-of-sight halos, then an over-correction of $H_0$ occurs such that it is biased low by $\sim~7$ per cent on average. These results were obtained from our simulations where we measure a relatively high median external convergence of $\kappa_{\rm ext} \simeq 0.11$ but we find that the size of the fractional bias in $H_0$ scales almost proportionally with $\kappa_{\rm ext} = 0.11$ on average (see below for details). Nevertheless, even with low levels of external convergence, this effect can not be ignored, since the uncertainties of current measurements of $H_0$ from strong lensing time delays are typically quoted as being lower than this \citep{Bonvin2017, Chen2019, Wong2019, Birrer2019, Rusu2020}. With the forthcoming large sample of strong lensing time delay systems observed by the future time domain large scale surveys, e.g., Mephisto\footnote{\url{http://www.swifar.ynu.edu.cn/info/1015/1073.htm}} and LSST, the effect becomes even more problematic.

Qualitatively, our conclusions are consistent with those of \cite{McCully2017} in the sense that line-of-sight structures significantly affect the accuracy of the measurement of $H_0$. We find a larger median external convergence of $\kappa_{\rm ext} \simeq 0.11$ compared to the value of 0.05 from \cite{McCully2017}. We attribute this to the fact that we have selected central galaxies of galaxy groups as the primary lenses in our light cones and because we have included more line-of-sight structures; we include galaxies from cosmoDC2 down to an $r$-band apparent magnitude of $28$, compared to the $i$-band limit of $21.5$ adopted by \cite{McCully2017}. Nevertheless, our findings indicate that even small values of $\kappa_{\rm ext}$ bias $H_0$ on average. We have shown that the fractional bias in inferred $H_0$ correlates with median external convergence according to the linear relationship $\Delta H_0 / H_0 = 0.626 \kappa_{\rm ext} - 0.005$.

We have also investigated the effects of incorporating external shear in the lens model. In the simulations using line-of-sight halos, adding an external shear term to the $SEPL$ lens model makes a negligible impact on the distribution of recovered values of $H_0$. Not unexpectedly, we also find that correcting this $SEPL$+$\gamma_{\rm ext}$ model with the average constant external convergence also leads to a $\sim 7$ per cent underestimation, which implies that the influence of external shear is negligible in the case of our study. This conclusion differs from that of McCully et al., most likely because we cleaned our lens sample by removing secondary halos that give rise to an Einstein radius of greater than 0.3 arcsec.

The Outer Rim simulations used to populate our lensing light cones with halos include only dark matter. As such, we have used SIE profiles in place of identified halos to better represent the total mass (baryons + dark matter) profiles of real lens galaxies. One effect this may have is that the lensing strength of any lower mass halos, which in the real Universe may not have accrued baryons, could be artificially enhanced by the more efficient isothermal profile. In addition, our simulated datasets do not include any large scale structure such as filaments although this is expected to be a small effect. We have explored the use of truncated SIE profiles in place of the non-truncated profiles used in this work but find that our results do not change significantly. Finally, we have ignored the effects of environmental structure in the simulations in the sense that our assumed smooth SIE profiles for the primary lens do not include substructure. We will leave consideration of these additional effects for future work.

To summarise, simple corrections for line-of-sight structure such as external shear or external convergence in estimations of $H_0$ using lensed time delays can not be relied upon in general. Time delay studies opt for lens systems that are apparently free of strong perturbers in an attempt to exclude line-of-sight effects, or they select systems where the perturbers are low in number and can be easily incorporated in the lens model. Our simulations have mimicked the former selection to a degree by removing halos from all of our light cones that produce a deflection resulting in an Einstein radius larger than 0.3 arcsec. Since this may still allow a significant flexion shift, an improved technique is to include perturbers in the lens model with a flexion shift above a certain threshold \cite[e.g.,][]{Rusu2020}. However, our work reveals that the culmination of many small line of sight perturbers continues to result in a significant portion of the measured bias in $H_0$ and more sophisticated modelling methods, for example, including more lens planes by lowering flexion shift thresholds are key to reliable measurements of $H_0$ from the hundreds of well-measured time-delay systems anticipated in forthcoming large strong lens samples.

\section*{Acknowledgements}
The authors thank the referee for instructive comments and suggestions to improve the manuscript. The authors are also thankful to Sherry Suyu and Thomas Collett for inspiring discussion and advice. We are grateful to Charles R. Keeton and Masamune Oguri for taking their time to answer our questions. We are incredibly thankful to Simon Birrer for making it possible to use \textsc{lenstronomy} for this project.NL and CB acknowledges support by the UK Science and Technology Facilities Council (STFC). SD is supported by the UK's STFC Ernest Rutherford Fellowship scheme. This research made use of CosmoDC2\footnote{\url{https://portal.nersc.gov/project/lsst/cosmoDC2}} and GCR-Catalogs-Reader\footnote{\url{https://github.com/LSSTDESC/gcr-catalogs}} created by the LSST Dark Energy Science Collaboration (DESC).
This work used the DiRAC@Durham facility managed by the Institute for Computational Cosmology on behalf of the STFC DiRAC HPC Facility (www.dirac.ac.uk). The equipment was funded by BEIS capital funding via STFC capital grants ST/K00042X/1, ST/P002293/1, ST/R002371/1 and ST/S002502/1, Durham University and STFC operations grant ST/R000832/1. DiRAC is part of the National e-Infrastructure.


\bibliographystyle{mnras}
\bibliography{ms} 


\section*{Data availability}
The data underlying this article will be shared on reasonable request to the corresponding author.


\label{lastpage}
\end{document}